\documentclass[10pt]{article}
\usepackage[dvips]{graphicx}
\begin{document}
\begin{center}
\textbf{{\large Dilaton Coupled Quintessence Model in the
$\omega-\omega'$ Plane}}
 \vskip 0.35 in
\begin{minipage}{4.5 in}
\begin{center}
{\small Z. G. HUANG$^\dag$ \vskip 0.06 in \textit{
Department~of~Mathematics~and~Physics,
\\~Huaihai~Institute~of~Technology,~222005,~Lianyungang,~China
\\
$^\ddag$zghuang@hhit.edu.cn}} \vskip 0.25 in {\small H. Q.
LU$^\ddag$ and W. FANG \vskip 0.06 in \textit{
Department~of~Physics,~Shanghai~University,~Shanghai,~China
\\
$^\dag$alberthq$\_$lu@staff.shu.edu.cn}}
\end{center}
\vskip 0.2 in

{\small In this paper, we regard dilaton in Weyl-scaled induced
gravitational theory as a coupled quintessence. Based on this
consideration, we investigate the dilaton coupled quintessence(DCQ)
model in $\omega-\omega'$ plane, which is defined by the equation of
state parameter for the dark energy and its derivative with respect
to $N$(the logarithm of the scale factor $a$). We find the scalar
field equation of motion in $\omega-\omega'$ plane, and show
mathematically the property of attractor solutions which correspond
to $\omega_\sigma\sim-1$, $\Omega_\sigma=1$. Finally, we find that
our model is a tracking one which belongs to "freezing" type model
classified in $\omega-\omega'$ plane. \vskip 0.2 in
\textit{Keywords:} Dark energy; Dilaton; Coupled Quintessence;
$\omega-\omega'$ plane; Attractor.
}
\end{minipage}
\end{center}
\vskip 0.2 in
\begin{flushleft}\textbf{1. Introduction}\end{flushleft}
Recent observations of high-redshift Type Ia Supernova[1] and the
Cosmic Microwave Background[2] have shown us such a fact: the
density of clustered matter including cold dark matters plus
baryons, $\Omega_{m0}\sim1/3$, and that the Universe is flat to high
precision, $\Omega_{total}=0.99\pm0.03$[3]. That is to say, we are
living in a flat universe which it undergoing a phase of accelerated
expansion, and there exists an unclumped form of energy density
pervading the Universe. This unknown energy density which is called
"dark energy" with negative pressure, contributes to two thirds of
the total energy density.
\par Perhaps the simplest explanation for
these data is that the dark energy corresponds to a positive
cosmological constant. However, the cosmological constant model
suffers from two serious issues called "coincidence problem" and
"fine-tuning problem". An alternative is a scalar field which has
not yet reached its ground state. These scalar field models include
Quintessence[4-14], K-essence[15], Tachyon[16], Phantom[17-20],
Quintom[21] and so on. The essential characteristics of these dark
energy models are contained in the parameter of its equation of
state, $p=\omega\rho$, where $p$ and $\rho$ denote the pressure and
energy density of dark energy, respectively, and $\omega$ is a state
parameter. Quintessence model has been widely studied, and its state
parameter $\omega$ which is time-dependent, is greater than $-1$. In
this paper, we regard dilaton in Weyl-scaled induced gravitational
theory as a coupled Quintessence. We call our model Dilaton Coupled
Quintessence(DCQ) model. Motivations that make us consider DCQ are
as follows: First, the dilaton is an essential element of string
theories and the low-energy string effective action[22]. Second,
dimensional compaction of Kaluza-Klein theories may naturally lead
to the Weyl-scaled induced gravitational theory. Third , dilatonic
gravities are expected to have such important cosmological
applications as in the case of (hyper)extended inflation[23].
Fourth, Many authors have considered the coupled dark energy[24]. In
our previous paper[25], we have considered a dilatonic dark energy
model, based on Weyl-scaled induced gravitational theory. We found
that when the dilaton field was not gravitational clustered at small
scales, the effect of dilaton can not change the evolutionary law of
baryon density perturbation, and the density perturbation can grow
from $z\sim10^3$ to $z\sim5$, which guarantees the structure
formation. When dilaton energy is very small compared the matter
energy, potential energy of dilaton field can be neglected. In this
case, the solution of cosmological scale $a$ has been found[26]. In
another paper[27], we have investigated the property of the
attractor solutions and concluded that the coupling between dilaton
and matter affects the evolutive process of the Universe, but not
the fate of the Universe.
\par Recently, many authors have investigated the evolution
of Quintessence dark energy models in the $\omega-\omega'$
plane[28,29], where $\omega'$ is the time variation of $\omega$ with
respect to $N$. According to different regions in the
$\omega-\omega'$ phase plane, these models can be classified two
types which are call "thawing" and "freezing" models. In what
follows, we shall study the cosmological dynamics of DCQ model in
the $\omega-\omega'$ plane. In the exponential potential
$Ae^{-\beta\sigma}$, the evolutive behaviors of dynamics of DCQ
model in the $\omega-\omega'$ plane, the energy density parameter of
dark energy $\Omega_\sigma$ with respect to $N$ are shown
mathematically. Our results show that the critical point with
$\omega\sim-1$ is the late-time attractor, where dilaton field
becomes ultimately frozen, as shown in the Fig.1. The evolution of
$\Omega_\sigma$ shows also that there exists a late-time attractor
solution, which corresponds to $\Omega_\sigma=1$.
\par Now let us consider the action of the Weyl-scaled induced gravitational theory:
\begin{equation}S=\int{d^4X\sqrt{-g}[\frac{1}{2}R(g_{\mu\nu})-\frac{1}{2}g^{\mu\nu}\partial_\mu\sigma\partial_\nu\sigma-W(\sigma)+L_{fluid}(\psi)}]\end{equation}
where
$L_{fluid}(\psi)=\frac{1}{2}g^{\mu\nu}e^{-\alpha\sigma}\partial_\mu\psi\partial_\nu\psi-e^{-2\alpha\sigma}V(\psi)$,
$\alpha=\sqrt{\frac{\kappa^2}{2\varpi+3}}$ with $\varpi>3500$[30]
being an important parameter in Weyl-scaled induced gravitational
theory, $\sigma$ is DCQ field, $g_{\mu\nu}$ is the Pauli metric
which can really represent the massless spin-two graviton and should
be considered to be physical metric[31]. We work in
units($\kappa^2\equiv8\pi G=1$). From the solar system tests, the
current constrain is $\alpha^2<0.001$[32]. The new constrain on the
parameter is $\alpha^2<0.0001$[33], which seems to argue against the
existence of long-range scalars. Perhaps such a pessimistic
interpretation of the limit is premature [31,32]. The conventional
Einstein gravity limit occurs as $\sigma\rightarrow 0$ for an
arbitrary $\varpi$ or $\varpi\rightarrow\infty$ with an arbitrary
$\sigma$. When $W(\sigma)=0$, it will result in the
Einstein-Brans-Dicke theory.
\par By varying action(1) and working in FRW universe, we obtain the field
equations of Weyl-scaled induced gravitational theory:
\begin{equation}H^2=\frac{1}{3}[\frac{1}{2}\dot{\sigma}^2+W(\sigma)+e^{-\alpha\sigma}\rho]\end{equation}
\begin{equation}\ddot{\sigma}+3H\dot{\sigma}+\frac{dW}{d\sigma}=\frac{1}{2}\alpha e^{-\alpha\sigma}(\rho-3p)\end{equation}
\begin{equation}\dot{\rho}+3H(\rho+p)=\frac{1}{2}\alpha\dot{\sigma}(\rho+3p)\end{equation}
where $H$ is Hubble parameter. For radiation $\rho_r=3p_r$, we get
$\rho_r\propto \frac{e^{\alpha\sigma}}{a^4}$ from Eq.(4). For matter
$p_m=0$, we get $\rho_m\propto
\frac{e^{\frac{1}{2}\alpha\sigma}}{a^3}$ from Eq.(4). Taking these
results into Eqs.(2), we obtain
\begin{equation}H^2=H_i^2[\frac{\frac{1}{2}\dot{\sigma}^2+W(\sigma)}{\rho_{c,i}}+\Omega_{m,i}e^{-\frac{1}{2}\alpha\sigma}(\frac{a_i}{a})^3+\Omega_{r,i}(\frac{a_i}{a})^4]\end{equation}
where $H_i^2=\frac{\rho_{c,i}}{3}$, $\rho_{c,i}$ is the critical
energy density of the universe at initial time $t_i$. $H_i$,
$\Omega_{m,i}$, $\Omega_{r,i}$ denote the Hubble parameter, matter
energy density parameter, radiation energy density parameter at
initial time $t_i$ respectively. We define our starting point as the
equipartition epoch, at which $\Omega_{m,i}=\Omega_{r,i}=0.5$ and
consider the initial scale factor $a_i=1$ for convenience. According
to the transformation $N=lna$, we have
\begin{equation}H=H_i[\frac{\frac{1}{2}\dot{\sigma}^2+W(\sigma)}{\rho_{c,i}}+\Omega_{m,i}e^{-\frac{1}{2}\alpha\sigma}e^{-3N}+\Omega_{r,i}e^{-4N}]^{\frac{1}{2}}\end{equation}
\vskip 0.2 in  \begin{flushleft}\textbf{2. Dynamics of DCQ Model in
the $\omega-\omega'$ Plane}\end{flushleft} The effective density
$\rho_{\sigma}$ and effective pressure $p_{\sigma}$ can be expressed
as follows
\begin{equation}\rho_{\sigma}=\frac{1}{2}\dot{\sigma}^2+W(\sigma)\end{equation}
\begin{equation}p_{\sigma}\equiv\omega_\sigma\rho_\sigma=\frac{1}{2}\dot{\sigma}^2-W(\sigma)\end{equation}
Substituting Eqs.(7)(8) into Eq.(4), we get
\begin{equation}-\frac{\dot{\rho}_\sigma}{H}=3\dot{\sigma}^2-\frac{\alpha\dot{\sigma}[\dot{\sigma}^2-W(\sigma)]}{H}\end{equation}
Because $\frac{\dot{\rho}_\sigma}{H}=\frac{d\rho_\sigma}{dN}$,
$\dot{\sigma}=\sqrt{2[\rho_\sigma-W(\sigma)]}$ and
$W(\sigma)=\frac{1}{2}(1-\omega_\sigma)\rho_\sigma$, the above
equation becomes
\begin{equation}-\frac{d\rho_\sigma}{dN}\equiv n_\sigma\rho_\sigma=\rho_\sigma[3(1+\omega_\sigma)-\frac{\alpha(1+3\omega_\sigma)}{2}\times\frac{\sqrt{3(1+\omega_\sigma)\rho_\sigma}}{H}]\end{equation}
Eq.(10) is the continuity equation of dilaton scalar field in DCQ
model. The evolutive equation of dilaton field can be expressed as
follows
\begin{equation}\frac{d\sigma}{dN}=\frac{\dot{\sigma}}{H}=\frac{\sqrt{6[\rho_\sigma-W(\sigma)]}}{H}=\frac{\dot{\sigma}}{H}=\frac{\sqrt{3(1+\omega_\sigma)\rho_\sigma}}{H}\end{equation}
Now, we define a new function
\begin{equation}\Delta(a)\equiv \frac{d(lnW(\sigma))}{d(ln\rho_\sigma)}=1+\frac{1}{1-\omega_\sigma}\times\frac{-d\omega_\sigma/dN}{\frac{1}{\rho_\sigma}(d\rho_\sigma/dN)}\end{equation}
So, we can rewrite Eq.(10) as
\begin{equation}\frac{d\omega_\sigma}{dN}=[3(1-\omega^2_\sigma)-\alpha\frac{(1-\omega_\sigma)(1+3\omega_\sigma)}{2}\ast\frac{\sqrt{3\rho_\sigma(1+\omega_\sigma)}}{H}]\times[\Delta-1]\end{equation}
where
\begin{equation}\Delta=\frac{\pm W'(\sigma)/W(\sigma)}{\frac{\sqrt{3(1+\omega_\sigma)\rho_\sigma}}{H}-\alpha\frac{1+3\omega_\sigma}{2}}\end{equation}
and the sign $"'"$ denotes the derivative of $W(\sigma)$ with
respect to $\sigma$. Eq.(13) is the scalar field equation of motion
in DCQ model. For the DCQ field rolling down its potential, the
$\pm$ sign before $W'(\sigma)$ corresponds to $W'(\sigma)>0$ or
$W'(\sigma)<0$. The size of the parameter
$\alpha=\sqrt{\frac{1}{2\varpi+3}}$ denotes the coupling intensity.
\par For a quintessence field without coupling to matter, the scalar
equation of motion is
\begin{equation}\frac{d\omega_\sigma}{dN}=[3(1-\omega^2_\sigma)]\times[\Delta^\ast-1]\end{equation}
where $\Delta^\ast=\frac{\pm
W'(\sigma)/W(\sigma)}{\frac{\sqrt{3(1+\omega_\sigma)\rho_\sigma}}{H}}$.
Obviously, when $\varpi\rightarrow\infty$ i.e. $\alpha\rightarrow0$,
Eq.(13) reduces to Eq.(15).
\par Now we consider the exponential
potential $W(\sigma)=Ae^{-\beta\sigma}$. The system decided by
Eq.(13) admits these critical points $\omega_\sigma=1$,
$\omega_\sigma\sim-1$, $\Delta=1$. When $\Delta=1$, $\omega_\sigma$
varies very slowly where the DCQ field is tracking and the ratio of
kinetic energy to potential energy of the DCQ field becomes a
constant. When $\omega_\sigma\sim-1$, it corresponds to a late time
attractor where the DCQ field becomes ultimately frozen, as shown in
the Fig.1. When we set different $\alpha$ value, the critical point
always tends $\omega_\sigma\sim-1$, as shown in Figs1-3. Fig.4 shows
the evolution of the DCQ fraction of the total energy density
$\Omega_\sigma$ from 0 to 1 with respect to $N$. Since
$T_{eq}\simeq5.64(\Omega_0h^2)eV\simeq2.843\times10^4K$,
$T_0\simeq2.7K$, $a_i=1$, the scale factor at the present epoch
$a_0$ would nearly be $1.053\times10^4$, then we know
$N_0=lna_0=9.262$. According to $N_0$, we obtain the current value
$\Omega_\sigma\simeq0.713021$, which meets the current observations
well.

\vskip 0.15 in
\begin{center}
\begin{minipage}{0.6\textwidth}
\includegraphics[scale=1.0]{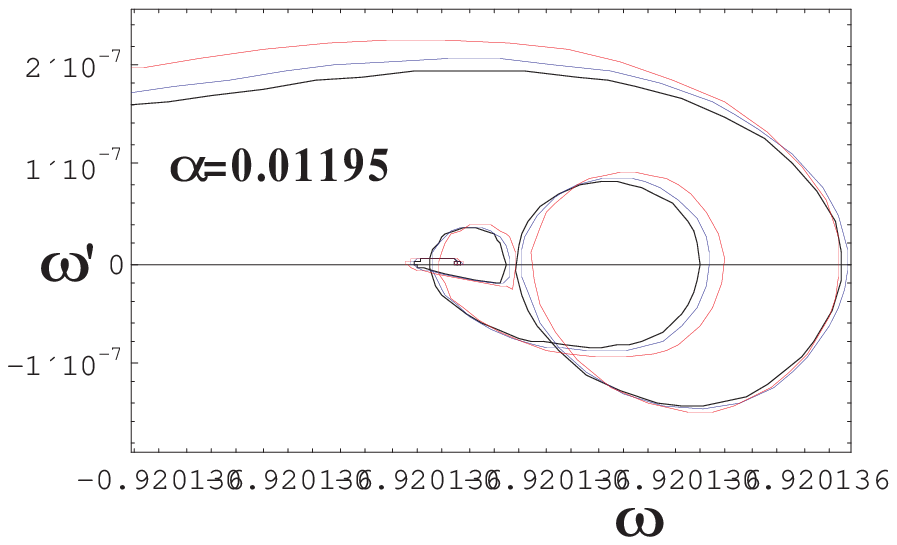}
{\small Fig.1 The comparison of attractive property of DCQ model
with exponential potential $W(\sigma)=Ae^{-\beta\sigma}$ for
different initial $\sigma_0=0.1$, 2 and 4, in $\omega-\omega'$
plane. We set $\alpha=0.01195$, $\beta=0.5$ and $A=1.0$.}
\end{minipage}
\end{center}

\vskip 0.3 in
\begin{center}
\begin{minipage}{0.6\textwidth}
\includegraphics[scale=1.0]{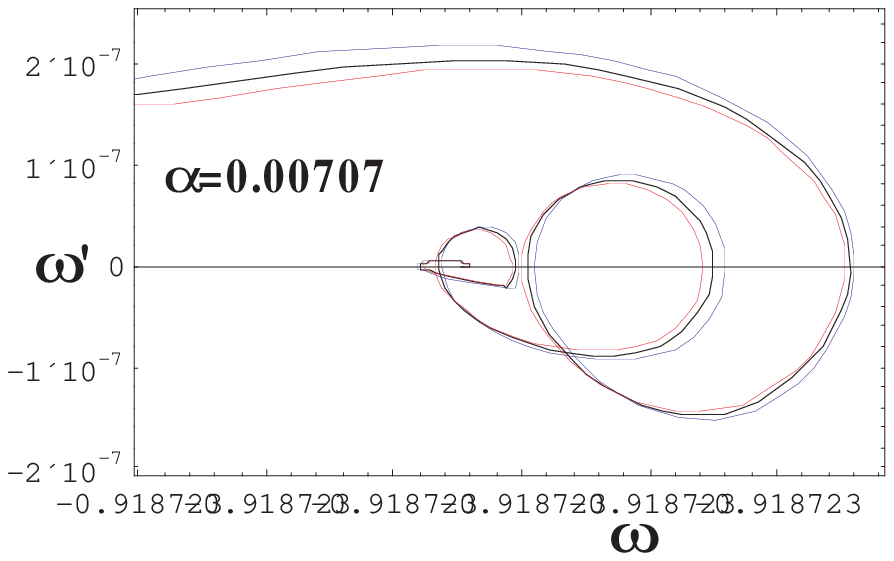}
{\small Fig.2 The comparison of attractive property of DCQ model
with exponential potential $W(\sigma)=Ae^{-\beta\sigma}$ when
$\alpha=0.00707$, $\beta=0.5$ and $A=1.0$.}
\end{minipage}
\end{center}

\vskip 0.3 in
\begin{center}
\begin{minipage}{0.6\textwidth}
\includegraphics[scale=1.0]{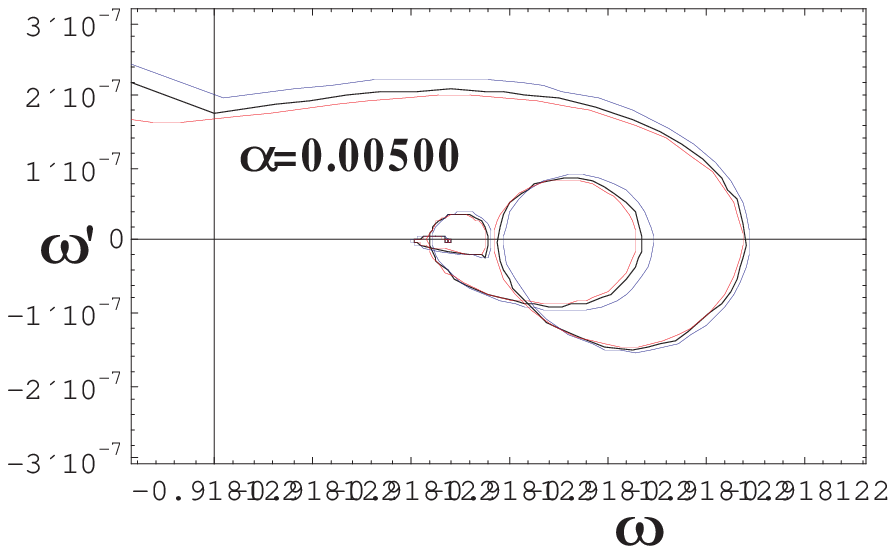}
{\small Fig.3 The comparison of attractive property of DCQ model
with exponential potential $W(\sigma)=Ae^{-\beta\sigma}$ when
$\alpha=0.00500$, $\beta=0.5$ and $A=1.0$.}
\end{minipage}
\end{center}

\vskip 0.3 in
\begin{center}
\begin{minipage}{0.6\textwidth}
\includegraphics[scale=1.0]{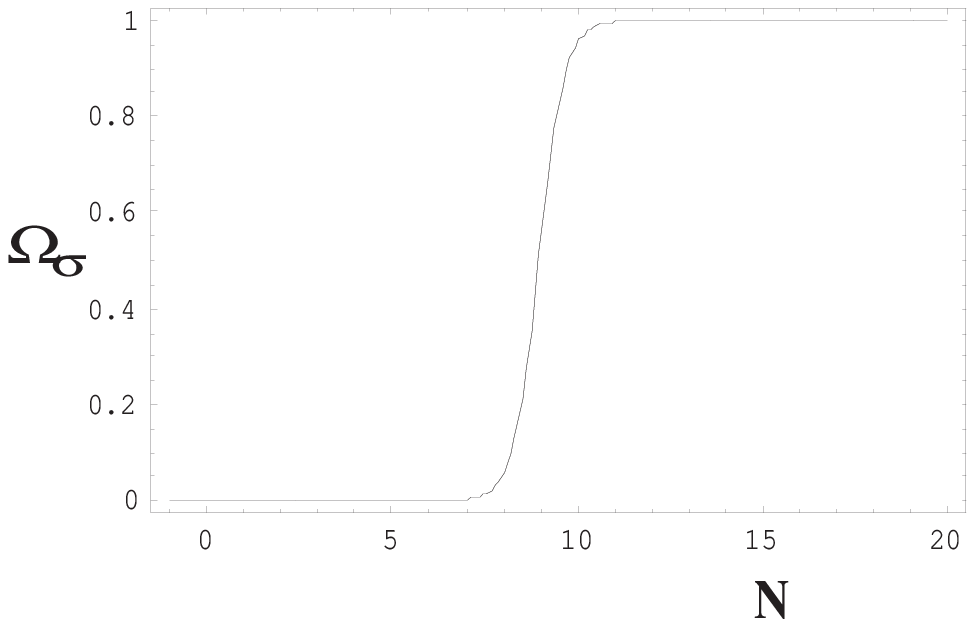}
{\small Fig.2 The evolution of the DCQ fraction of the total energy
density $\Omega_\sigma$ with respect to N in exponential potential
$W(\sigma)=Ae^{-\beta\sigma}$. We set $\alpha=0.01195$, $\beta=0.5$
and $A=1.0$. Since the current value $N_0=lna_0=9.262$, we obtain
the current value $\Omega_\sigma\simeq0.713021$, which meets the
current observations well.}
\end{minipage}
\end{center}

\vskip 0.3 in
\begin{flushleft}\textbf{{3. Conclusions}}\end{flushleft}
\par Caldwell and Linder classified the quintessence models into two
types "thawing" and "freezing" model according to the different
regions in $\omega-\omega'$ plane and gave the limit of
quintessence[28]: $1+\omega\geq0.004$ for "thawing" model and
$1+\omega\geq0.01$ for "freezing" model. In DCQ model, we can also
make the same delimitation according to different initial
conditions. For the "thawing" type model, the DCQ field has been
frozen by Hubble damping at a value displace from its minimum until
recently, when it starts to roll down to the minimum. For the
"freezing" model, the DCQ field which was already rolling down
towards its potential minimum, prior to the onset of acceleration,
but which slows down and creeps to a halt as it comes to dominate
the Universe. We find our DCQ model belongs to the "freezing" type
model and satisfies the limit of "freezing" model:
$1+\omega_\sigma=1+(-0.920136)=0.079864>0.01$. We investigate the
cosmological dynamics of DCQ model with exponential potential
$Ae^{-\beta\sigma}$ in the $\omega-\omega'$ plane and examine the
energy density parameter of dark energy $\Omega_\sigma$ with respect
to $N$. These numerical results show that the critical point with
$\omega\sim-1$ is the late-time attractor, where dilaton field
becomes ultimately frozen and $\Omega_\sigma=1$. We also consider
the effect of the parameter $\alpha$ on the property of attractor
solution. When we set $\alpha=0.01195,~0.00707,~0.00500$, the
evolutive behavior of DCQ attractor changes so tiny that it can be
neglected. This result is consonant with our previous viewpoint that
the coupling between dilaton and matter affects the evolutive
process of the Universe, but not the fate of the Universe[27].
According to $N_0=ln a_0\doteq9.262$, we obtain the current value of
$\Omega_\sigma$ is 0.713021, which meets the current observations
well.
\begin{flushleft}\textbf{Acknowledgements}\end{flushleft}
This work is partially supported by National Nature Science
Foundation of China under Grant No.10573012 and Shanghai Municipal
Science and Technology Commission No.04dz05905.

\begin{flushleft}{\noindent\bf References}
 \small{

\item {1.}{ A. G. Riess et al., \textit{Astrophys. J}\textbf{607}, 665(2004);
\\\hspace{0.15 in}A. G. Riess, \textit{Astron. J}\textbf{116}, 1009(1998);
\\\hspace{0.15 in}S. Perlmutter et al., \textit{Astrophys. J}\textbf{517}, 565(1999);
\\\hspace{0.15 in}N. A. Bahcall et al., \textit{Science}\textbf{284}, 1481(1999).}
\item {2.}{ D. N. Spergel et al., Astrophys. J. Suppl\textbf{148}, 175(2003).}
\item {3.}{ P. de Bernardis et al., arXiv:astro-ph/0105296;
\\\hspace{0.15 in}R. Stompor et al., arXiv:astro-ph/0105062;
\\\hspace{0.15 in}C. Pryke et al., arXiv:astro-ph/0104490.}
\item {4.}{ J. S. Bagla, H. K. Jassal and T. Padmamabhan, \textit{Phys. Rev. D}\textbf{67}, 063504(2003).}
\item {5.}{ L. Amendola, M. Quartin, S. Tsujikawa and I. Waga, \textit{Phys.Rev.D}\textbf{74}, 023525(2006);
\\\hspace{0.15 in}E. Elizalde, S. Nojiri and S. D. Odintsov, arXiv:hep-th/0405034;
\\\hspace{0.15 in}S. Nojiri, S. D. Odintsov and M. Sasaki, \textit{Phys.Rev. D}\textbf{70} 043539(2004);
\\\hspace{0.15 in}S. Nojiri and S. D. Odintsov, arXiv:hep-th/0601213;
\\\hspace{0.15 in}S. Nojiri and S. D. Odintsov \textit{Phys. Lett. B}\textbf{639}, 144(2006);
\\\hspace{0.15 in}S. Nojiri, S. D. Odintsov and H. Stefancic, arXiv:hep-th/0608168;
\\\hspace{0.15 in}S. Nojiri and S. D. Odintsov, \textit{Phys. Lett. B}\textbf{562}, 147(2003)[arXiv:hep-th/0303117];
\\\hspace{0.15 in}S. Nojiri and S.D. Odintsov, \textit{Phys. Rev. D}\textbf{70}, 103522(2004)[arXiv:hep-th 0408170];
\\\hspace{0.15 in}S. Nojiri, S. D. Odintsov and S. Tsujikawa, arXiv:hep-th/0501025;
\\\hspace{0.15 in}S. Nojiri and S. D. Odintsov, \textit{Phys. Rev. D}\textbf{7}2, 023003(2005)[arXiv:hep-th/0505215];
\\\hspace{0.15 in}S. Nojiri and S. D. Odintsov, arXiv:hep-th 0506212;
\\\hspace{0.15 in}S. Nojiri and S. D. Odintsov, arXiv:hep-th/0611071;
\\\hspace{0.15 in}B. Boisseau et al., \textit{Phys. Rev. Lett}\textbf{85}, 2236(2000)[arXiv:gr-qc/0001066];
\\\hspace{0.15 in}G. Esposito-Farese and D. Polarski, \textit{Phys. Rev. D}\textbf{63}, 063504(2001)[arXiv:gr-qc/0009034];
\\\hspace{0.15 in}Xin Zhang, \textit{Mod. Phys. Lett. A}\textbf{20}, 2575(2005)[arXiv:astro-ph/0503072];
\\\hspace{0.15 in}Xin Zhang, \textit{Phys. Lett. B}\textbf{611}, 1(2005)[arXiv:astro-ph/0503075];
\\\hspace{0.15 in}M. R. Setare, \textit{Phys. Lett. B}\textbf{642},1(2006)[arXiv:hep-th/0609069];
\\\hspace{0.15 in}M. R  Setare, arXiv:hep-th/0609104;
\\\hspace{0.15 in}M. R. Setare, arXiv:hep-th/0610190.}
\item {6.}{ C. Wetterich \textit{Nucl. Phys. B}\textbf{302}, 668(1998);
\\\hspace{0.15 in}E. J. Copeland, M. Sami and S. Tsujikawa, arXiv:hep-th/0603057;
\\\hspace{0.15 in}P. G. Ferreira and M. Joyce \textit{Phys. Rev. D}\textbf{58}, 023503(1998);
\\\hspace{0.15 in}J. Frieman, C. T. Hill, A. Stebbinsand and I.Waga, \textit{Phys. Rev. Lett}\textbf{75}, 2077(1995);
\\\hspace{0.15 in}P. Brax and J. Martin, \textit{Phys. Rev. D}\textbf{61}, 103502(2000);
\\\hspace{0.15 in}T. Barreiro, E. J. Copeland and N. J. Nunes, \textit{Phys. Rev. D}\textbf{61}, 127301(2000);
\\\hspace{0.15 in}I. Zlatev, L. Wang and P. J. Steinhardt \textit{Phys. Rev. Lett} \textbf{82}, 896(1999).}
\item {7.}{ T. Padmanabhan, and T. R. Choudhury, \textit{Phys. Rev. D}\textbf{66}, 081301(2002).}
\item {8.}{ A. Sen, \textit{JHEP} \textbf{0204}, 048(2002).}
\item {9.}{ C. Armendariz-Picon, T. Damour and V. Mukhanov, \textit{Phys. Lett. B}\textbf{458}, 209(1999).}
\item {10.}{A. Feinstein, \textit{Phys. Rev. D}\textbf{66}, 063511(2002);
\\\hspace{0.17 in}M. Fairbairn and M. H. Tytgat, \textit{Phys. Lett. B}\textbf{546} 1(2002).}
\item {11.}{A. Frolov, L. Kofman and A. Starobinsky, \textit{Phys.Lett.B} \textbf{545}, 8(2002);
\\\hspace{0.17 in}L. Kofman and A. Linde, \textit{JHEP}\textbf{0207}, 004(2004).}
\item {12.}{C. Acatrinei and C. Sochichiu, \textit{Mod. Phys. Lett. A}\textbf{18}, 31(2003);
\\\hspace{0.17 in}S. H. Alexander, \textit{Phys. Rev. D}\textbf{65}, 0203507(2002).}
\item{13.}{T. Padmanabhan, \textit{Phys. Rev. D}\textbf{66}, 021301(2002).}
\item{14.}{A. Mazumadar, S. Panda and A. Perez-Lorenzana, \textit{Nucl. Phys. B}\textbf{614}, 101(2001);
\\\hspace{0.17 in}S. Sarangi and S. H. Tye, \textit{Phys. Lett. B}\textbf{536}, 185(2002).}
\item{15.}{C. Armend\'{a}riz-Pic\'{o}n, V. Mukhanov and P. J. Steinhardt, \textit{Phys.Rev.Lett}\textbf{85}, 4438(2000);
\\\hspace{0.17 in}C. Armend\'{a}riz-Pic\'{o}n, V. Mukhanov and P. J. Steinhardt, \textit{Phys. Rev. D}\textbf{63}, 103510(2001);
\\\hspace{0.17 in}T. Chiba, \textit{Phys.Rev.D}\textbf{66}, 063514(2002);
\\\hspace{0.17 in}M. Malquarti, E. J. Copeland, A. R. Liddle and M. Trodden, \textit{Phys. Rev. D}\textbf{67}, 123503(2003);
\\\hspace{0.17 in}R. J. Sherrer, \textit{Phys. Rev. Lett}\textbf{93)}, 011301(2004);
\\\hspace{0.17 in}L. P. Chimento, \textit{Phys. Rev. D}\textbf{69}, 123517(2004);
\\\hspace{0.17 in}A. Melchiorri, L. Mersini, C. J. Odman and M. Trodden, \textit{Phys. Rev. D}\textbf{68}, 043509(2003).}
\item{16.}{A. Sen, \textit{JHEP} \textbf{0207}, 065(2002);
\\\hspace{0.17 in}M. R. Garousi, \textit{Nucl. Phys. B}\textbf{584}, 284(2000);
\\\hspace{0.17 in}M. R. Garousi, \textit{JHEP} \textbf{0305}, 058(2003);
\\\hspace{0.17 in}E. A. Bergshoeff, M. de Roo, T. C. de Wit, E. Eyras and S. Panda,\textit{ JHEP} \textbf{0005}, 009(2000);
\\\hspace{0.17 in}J. Kluson, \textit{Phys. Rev. D}\textbf{62}, 126003(2000);
\\\hspace{0.17 in}G. W. Gibbons, \textit{Phys. Lett. B}\textbf{537}, 1(2002);
\\\hspace{0.17 in}M. Sami, P. Chingangbam and T. Qureshi, \textit{Phys. Rev. D}\textbf{66}, 043530(2002);
\\\hspace{0.17 in}M. Sami, \textit{Mod. Phys. Lett. A}\textbf{18}, 691(2003);
\\\hspace{0.17 in}Y. S. Piao, R. G. Cai, X. m. Zhang and Y. Z. Zhang, \textit{Phys. Rev. D}\textbf{66},121301(2002);
\\\hspace{0.17 in}L. Kofman and A. Linde, \textit{JHEP} \textbf{0207}, 004(2002).}
\item{17.}{H. Q. Lu, \textit{Int. J. Mod. Phys. D}\textbf{14}, 355(2005)[arXiv:hep-th/0312082];
\\\hspace{0.17 in}H. Q. Lu, Z. G. Huang, W. Fang and P. Y. Ji, arXiv:hep-th/0504038;
\\\hspace{0.17 in}W. Fang, H. Q. Lu, Z. G. Huang and K. F. Zhang, \textit{Int. J. Mod. Phys. D}\textbf{15}, 199(2006)[arXiv:hep-th/0409080];
\\\hspace{0.17 in}W. Fang, H. Q. Lu and Z. G. Huang, arXiv:hep-th/0606032.}
\item{18.}{X. Z. Li and J. G. Hao, \textit{Phys. Rev. D}\textbf{69}, 107303(2004).}
\item{19.}{T. Chiba, T. Okabe and M. Yamaguchi, \textit{Phys. Rev. D}\textbf{62}, 023511(2000);
\\\hspace{0.17 in}L. Amendola, S. Tsujikawa, and M. Sami, \textit{Phys. Lett. B}\textbf{632}, 155(2006);
\\\hspace{0.17 in}L. Amendola, \textit{Phys. Rev. Lett.}\textbf{93}, 181102(2004).}
\item{20.}{P. Singh, M. Sami and N. Dadhich, \textit{Phys.Rev. D}\textbf{68}, 023522(2003);
\\\hspace{0.17 in}S. M. Carroll, M. Hoffman and M. Trodden, \textit{Phys. Rev. D}\textbf{68}, 023509(2003);
\\\hspace{0.17 in}R. Gannouji, D. Polarski, A. Ranquet and A. A. Starobinsky, arXiv:astro-ph/0606287.}
\item{21.}{W. Hao, R. G. Cai and D. F. Zeng, \textit{Class.Quant.Grav}\textbf{22}, 3189(2005);
\\\hspace{0.17 in}Z. K. Guo, Y. S. Piao, X. M. Zhang, Y.Z. Zhang, \textit{Phys.Lett. B}\textbf{608}, 177(2005);
\\\hspace{0.17 in}B. Feng, arXiv:astro-ph/0602156.}
\item{22.}{Polchinski J, 1998 String Theory, Cambridge University Press, Cambridge.}
\item{23.}{Kolb E W, Salopek D and Turner M S 1990 \textit{Phys.Rev D} \textbf{42} 3925.}
\item{24.}{L. Amendola, C. Quercellini, \textit{Phys. Rev. D}\textbf{68}, 023514(2003);
\\\hspace{0.17 in}Michael Doran and J. J\"{a}chel, \textit{Phys. Rev. D}\textbf{66}, 043519(2002);
\\\hspace{0.17 in}N. J. Nunes and D. F. Mota, \textit{Mon. Not. Roy. Astron. Soc.}\textbf{368}, 751(2006);
\\\hspace{0.17 in}M. Manera, D. F. Mota, \textit{Mon. Not. Roy. Astron. Soc.}\textbf{371}, 1373(2006);
\\\hspace{0.17 in}Ishwaree P. Neupane, arXiv:hep-th/0602097;
\\\hspace{0.17 in}Ishwaree P. Neupane and Benedict M. N. Carter, \textit{JCAP} \textbf{0606}, 004(2006);
\\\hspace{0.17 in}Ishwaree P. Neupane and Benedict M. N. Carter, \textit{Phys. Lett. B}\textbf{638}, 94(2006);
\\\hspace{0.17 in}Federico Piazza and Shinji Tsujikawa, \textit{JCAP} \textbf{0407}, 004(2004);
\\\hspace{0.17 in}B. Gumjudpai, T. Naskar, M. Sami, and S. Tsujikawa, \textit{JCAP} \textbf{0506}, 007(2005).}
\item{25.}{H. Q. Lu, Z. G. Huang, W. Fang and K. F. Zhang, arXiv:hep-th/0409309.}
\item{26.}{H. Q. Lu and K. S. Cheng, \textit{Astrophysics and Space Science}\textbf{235}, 207(1996);
\\\hspace{0.17 in}Z. G. Huang, H. Q. Lu and P. P. Pan, \textit{Astrophysics and Space Science}\textbf{295}, 493(2005);
\\\hspace{0.17 in}Y. G. Gong, arXiv:gr-qc/9809015.}
\item{27.}{Z. G. Huang, H. Q. Lu and W. Fang, \textit{Class. Quant. Grav}\textbf{23}, 6215(2006)[arXiv:hep-th/0604160];
\\\hspace{0.17 in}Z. G. Huang and H. Q. Lu, \textit{Int. J. Mod. Phys. D}\textbf{15}, 1501(2006).}
\item{28.}{R. R. Caldwell and E. V. Linder, \textit{Phys. Rev. Lett.}\textbf{95}, 141301(2005).}
\item{29.}{S. A. Bludman and M. Roos, \textit{Phys.Rev. D}\textbf{65}, 043503(2002);
\\\hspace{0.17 in}Z. K. Guo, Y. S. Piao, X. M. Zhang and Y. Z. Zhang[arXiv:astro-ph/0608165].}
\item{30.}{C. M. Will, \textit{Living Rev. Rel.}\textbf{4}, 4(2001).}
\item{31.}{Y. M. Cho, \textit{Phys. Rev.Lett}\textbf{68}, 3133(1992).}
\item{32.}{T. Damour and K. Nordtvedt, \textit{Phys. Rev. Lett}\textbf{70}, 2217(1993).}
\item{33.}{B. Bertotti, L. Iess and P. Tortora, \textit{Nature}\textbf{425}, 374(2003).}}
\end{flushleft}
\end{document}